\documentstyle[prb,aps,epsfig,eqsecnum,amsmath,amssymb]{revtex}

\begin{document}
\draft
%--------------------------------------------------------
\twocolumn[\hsize\textwidth\columnwidth\hsize\csname@twocolumnfalse\endcsname
%--------------------------------------------------------
%\title{ Magnetization properties of 2-dimensional spin-1/2 square lattices}
 
\title{The magnetization process in the 2-dimensional $J_1-J_2$ model}
\author{A. Fledderjohann, K.-H. M{\"u}tter}
\address{Physics Department, University of Wuppertal, 42097 Wuppertal, Germany} 
%\date{\today}
\maketitle
%%%%%%%%%%%%%%%%%%%%%%%%%%%%%%%%%%%%%%%%
%
\begin{abstract}
%Spuare lattices of spin-1/2 Heisenberg spins with orthogonal couplings
%of unit strength and additional diagonal couplings ($\alpha$) ..
We study the $\alpha=J_2/J_1$-dependence of the magnetization 
process in the $J_1-J_2$ model on a square lattice with
frustrating couplings $J_2$ along the diagonals.
Perturbation expansions around $\alpha=J_2/J_1=0$ and $\alpha^{-1}=0$ 
yield an adequate
description of the magnetization curve in the antiferromagnetic and collinear
antiferromagnetic phase, respectively. The transition from one phase to the other
($0.5<\alpha<0.7$) leaves pronounced structures in the longitudinal and transverse
structure factors at ${\bf p}=(\pi,\pi)$ and ${\bf p}=(0,\pi)$.
\end{abstract}
%
%%%%%%%%%%%%%%%%%%%%%%%%%%%%%%%%%%%%%%%%
\pacs{75.10.-b, 75.10.Jm}
%------------------------------------------------------
\twocolumn]
%%%%%%%%%%%%%%%%%%%%%%%%%%%%%%%%%%%%%%%%%%%%%%%%%%%%%%%%%%%%%%%%%%%%%%%%
 
\section{Introduction}

Magnetization processes in quasi one-dimensional quantum spin systems
have been studied intensively during the last years. The Oshikawa,
Yamanaka and Affleck quantization rule \cite{oshikawa97a} predicts
possible plateaus in the magnetization curve at certain rational values
of the magnetization $M$, which can be derived either from the
geometry of the unit cell or a mapping onto a one-dimensional system
with modulated couplings.\cite{fl99}

In the latter formulation, the plateaus appear at those values where
the wave vector of the modulation coincides with the soft mode momenta
of the unperturbed system. The existence of soft modes (zero energy
excitations) is guaranteed by the Lieb Schultz Mattis theorem 
\cite{lieb61} for translation invariant, one-dimensional systems
with finite-range couplings. A rigorous extension of this important
theorem to two and three-dimensional systems does not exist to date.
For this reason, the Oshikawa, Yamanaka and Affleck quantization
rule cannot be applied in a straightforward way to predict possible
plateaus in the magnetization curves of two- and three-dimensional
systems.

Experimental results, however, just concern compounds with a higher
dimensional coupling structure as there are:

(A) $CsCuCl_3$ [Ref. \onlinecite{nojiri88}] Here, a plateau has been found 
at $M/M_s=1/3$ ($M_s=1/2$ is the saturating magnetization). An
explanation of this feature has been given in Ref. \onlinecite{honecker00}.
%This compound
%is suggested to be built up from coupled 3-leg ladders (with periodic
%boundary conditions along the rungs). An isolated three leg ladder
%\cite{honecker00} has indeed a magnetization plateau at $M=1/6$. 
%It looks that the
%coupling between the ladders does not change the plateau structure
%in the magnetization curve. \cite{honecker00}

(B) $NH_4CuCl_3$ [Ref. \onlinecite{shiramura98}] Here, magnetization 
plateaus have been found for $M/M_s=1/4,\,3/4$.
The compound is suggested to be built up as a two-dimensional
structure of interacting two leg zig-zag ladders.\cite{kolezhuk99} 
%Non-interacting
%two leg zig-zag ladders cannot explain the observed plateaus.

(C) $SrCu_2(BO_3)_2$ [Ref. \onlinecite{onizuka00}] Magnetization 
plateaus have been observed
for $M/M_s=1/3,\,1/4,\,1/8$. The compound is modelled by the
two-dimensional Shastry Sutherland model.
\cite{shastry81,miyahara99,muellerhartmann00}
An explanation of the plateaus at $M/M_s=1/4,1/3$ has been discussed
in Ref. \onlinecite{momoi00}.
%by Momoi and Totsuka. \cite{momoi00}

In this paper we are going to study the magnetization process in
a two-dimensional spin-1/2 Heisenberg model with Hamiltonian

\begin{eqnarray}
\label{h0}
H & = & J_1H_1+J_2H_2-BS_3({\bf p}={\bf 0})
\end{eqnarray}
with isotropic couplings
\begin{eqnarray}
H_l & = & \sum_{{\bf x},{\bf y}}J_l({\bf x},{\bf y}){\bf S}({\bf x})\cdot 
{\bf S}({\bf y})\,,
\quad\quad l=1,2
\end{eqnarray}
for nearest and next-to-nearest neighbour sites $x$ and $y$:
\begin{eqnarray}
J_1({\bf x},{\bf y}) & = & \delta_{{\bf y},{\bf x}+\hat{1}}+\delta_{{\bf y},
{\bf x}+\hat{2}}\\
J_2({\bf x},{\bf y}) & = & \delta_{{\bf y},{\bf x}+\hat{1}+\hat{2}}+\delta_{
{\bf y},{\bf x}+\hat{1}-\hat{2}}\,;
\end{eqnarray}
$\hat{1}$ and $\hat{2}$ denote lattice vectors in horizontal and
vertical directions. For convenience we choose the nearest neighbour
coupling to be one ($J_1=1$, i.e. $\alpha=J_2/J_1=J_2$) and use 
\begin{eqnarray}
S_j({\bf p}) & = & \sum_{{\bf x}} e^{i{\bf p}\cdot{\bf x}}S_j({\bf x})
\,,\quad\quad j=1,2,3
\end{eqnarray}
for the Fourier transform of the spin operators on the square lattice.

If we assume periodic boundary conditions
the Hamiltonian \eqref{h0} is translationally invariant. Moreover 
it commutes with the total spin operators ${\bf S}^2({\bf p}={\bf 0})$
and $S_3({\bf p}={\bf 0})$.
We will classify the eigenstates $|E,{\bf p},s,s_3\rangle$ by the eigenvalues:

\begin{tabular}{ll}
$E$ & of $H_1+\alpha H_2$\,,\\
${\bf p}=(p_1,p_2)$ & of the momentum operator\,,\\
$s(s+1)$ & of the squared total spin ${\bf S}^2({\bf p}={\bf 0})$\,,\\
$s_3$ & of the 3-component $S_3({\bf p}={\bf 0})$\,.
\end{tabular}

The magnetization curve $M=M(B)$ is computed from the energy differences
\begin{eqnarray}\label{b_m}
B(M=s/N,\alpha) & = & E(s+1,{\bf p}_{s+1},\alpha)-E(s,{\bf p}_s,\alpha)\,.\nonumber\\
 & & 
\end{eqnarray}
$E(s,{\bf p}_s,\alpha)$ is the lowest eigenvalue of $H_1+\alpha H_2$
in the sector with total spin $s$. $N$ is the total number of sites
and ${\bf p}_s$ is the ground state momentum
in this sector.

${\bf p}_s$ can be deduced from Marshall's sign rule \cite{marshall55}
in the limiting cases $\alpha=0$ and $\alpha\rightarrow\infty$.

$a)$ $\alpha=0$

\begin{eqnarray}
{\bf p}_s=(0,0) & & \mbox{\quad if}\quad \frac{N}{2}+s\quad\mbox{is even}\\
{\bf p}_s=(\pi,\pi) & & \mbox{\quad if}\quad \frac{N}{2}+s\quad\mbox{is odd}\,,
\end{eqnarray}
where the latter equations hold for clusters with size $N=L\times L$,
$L$ even and periodic boundary conditions.
Therefore, the transition between two subsequent ground states as they
enter on the left- and right-hand side of \eqref{b_m} is accompanied by
a momentum transfer
\begin{eqnarray}\label{pi_pi}
{\bf q}={\bf p}_{s+1}-{\bf p}_s & = & (\pi,\pi)
\end{eqnarray}

$b)$ $\alpha\rightarrow\infty$

The Hamiltonian
\begin{eqnarray}
H_2 & = & H_1^{(+)}+H_1^{(-)}\label{h+-}
\end{eqnarray}
decays into two {\em nearest}-neighbour Hamiltonians on the even and
odd sublattices, respectively.
The nearest-neighbour couplings are
defined here along the diagonals $\hat{1}+\hat{2}$, $\hat{1}-\hat{2}$.
Seen from the original lattice, the diagonals are rotated by
$\pm\frac{\pi}{4}$. Moreover, the lattice constant increases by
a factor $\sqrt{2}$. 

We conclude from this that the momentum
transfer between the two subsequent ground states in \eqref{b_m}
is here
\begin{eqnarray}\label{pi_0}
{\bf q}={\bf p}_{s+1}-{\bf p}_s & = & (\pi,0),\,\,(0,\pi)\,.
\end{eqnarray}
The change from \eqref{pi_pi} to \eqref{pi_0} signals a phase transition
from antiferromagnetic to collinear antiferromagnetic order. \cite{schulz92}
%The transition will occur at some critical value $\alpha=\alpha_0(M)$
%which depends on the magnetization $M=s/N$ and thereby on the
%applied field.

Linear spin wave theory predicts that the regimes for antiferromagnetic
and collinear antiferromagnetic order are restricted to $\alpha<0.4$
and $\alpha>0.55$, respectively. In between a phase with transverse
disorder has been suggested. \cite{zhitomirsky00}

We pursue the following strategy to study the impact of frustration
on ground state energies and magnetization curves: We derive 
perturbation expansions in $\alpha$ and $\alpha^{-1}$, which are 
aimed to describe adequately the behaviour in the antiferromagnetic
and collinear antiferromagnetic phase, respectively.
Comparison with computations on finite clusters indicates that the
perturbation results agree for $\alpha<0.5$ and $\alpha>0.7$,
respectively. The regime inbetween which is not accessible by
perturbation methods is of special interest. Here,
we expect the emergence of plateaus in the magnetization curve.
\cite{zhitomirsky00,lozovik93}

%\newpage

The outline of the paper is as follows:
In Sec. II and Appendices \ref{strong_al} and \ref{2nd_order} 
we present the results for the ground state energies
obtained from perturbation expansions around $\alpha=0$ and
$\beta=1/\alpha=0$.

Free energies and magnetization curves are discussed in Sec. III.
Numerical results on the lowest frequency moments of the dynamical
structure factor for momenta ${\bf q}=(\pi,\pi)$ and ${\bf q}=
(\pi,0),\,(0,\pi)$ are given in Sec. IV.

%The approximate value of $\alpha_0(M)$ is already obtained by two
%perturbation expansions around $\alpha=0$ and $\beta=1/\alpha=0$.

%%%%%%%%%%%%%%%%%%%%%%%%%%%%%%%%%%%%%%%%%%%%%%%%%%%%%%%%%%%%%%%%%%%%%%%%

\section{From antiferromagnetic to collinear antiferromagnetic order}
\label{af_coll}

In the antiferromagnetic phase $\alpha<\alpha_0(M)$ perturbation
theory up to second order yields for the lowest eigenvalue of 
$H_1+\alpha H_2$ in the sector with magnetization $M=s/N$:

\begin{eqnarray}
E(s,{\bf p}_s,\alpha) & = & N\left( \epsilon_1(M)+\alpha\epsilon_2(M)+
\alpha^2\delta_2(M)\right)\,;\nonumber\\
 & & \quad\quad\quad\quad\quad\quad \alpha<\alpha_0(M)\label{2nd_lt}
\end{eqnarray}
where
\begin{eqnarray}\label{epsilon}
\epsilon_j(M) & = & \frac{1}{N}\langle 0|H_j|0\rangle\,,\quad\quad j=1,2
\end{eqnarray}
are the expectation values of $H_1$ and $H_2$ determined for the
unfrustrated ground states $|0\rangle$ (i.e. $H=H_1$).

The second order contribution $\delta_2(M)$ can be expressed in terms
of the transition probabilities $|\langle n|H_2|0\rangle|^2$ and energy differences
$E_n-E_0$ between the ground state $|0\rangle$ and the excited states 
$|n\rangle$:

\begin{eqnarray}
\delta_2(M) & = & -\frac{1}{N}\sum_n\frac{|\langle n|H_2|0\rangle|^2}{E_n-E_0}\,.
\end{eqnarray}

\begin{figure}[hb!]
\centerline{\hspace{-3mm}\epsfig{file=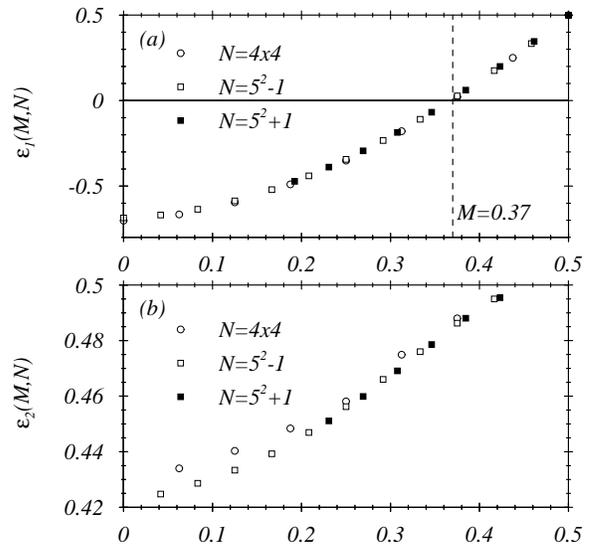,width=7.6cm,angle=0}}
\caption{$M$-dependence of $\epsilon_{1,2}(M)$ for system sizes $N=4\times 4$
(square lattice) and $N=5^2\pm1$ (helical boundary conditions)}
\label{fig2}
\end{figure}

The $M$-dependence of $\epsilon_1(M)$ and $\epsilon_2(M)$ is shown in
Fig.~\ref{fig2}(a) and \ref{fig2}(b) respectively for system sizes 
$N=4\times 4$ (with periodic
boundary conditions) and $N=5\times 5\mp 1=24,\,26$ (with helical
boundary conditions). The data points for $\epsilon_1(M)$ nicely scale
in $M$; deviations from scaling appear in $\epsilon_2(M)$ for smaller
$M$ values.

\begin{figure}[]
\centerline{\hspace*{0mm}\epsfig{file=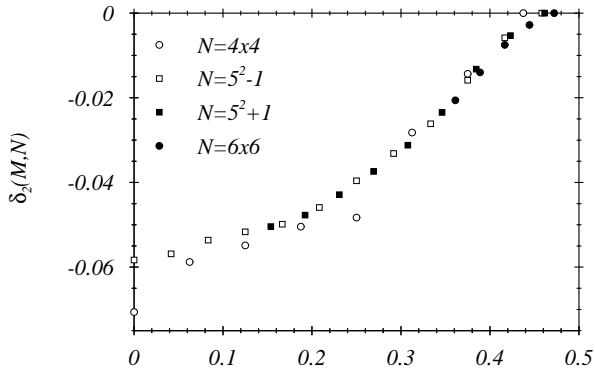,width=7.6cm,angle=0}}
\caption{$M$- and $N$-dependence of the second order contribution
$\delta_2(M)$}
\label{fig3}
\end{figure}

Numerical results for the second order contribution $\delta_2(M)$ obtained
with the recursion method  \cite{recursion} are shown in Fig.~\ref{fig3}. 
Scaling
in $M$ is realized for larger $M$-values ($M>0.3$). For $M=0$ and $M=1/4$,
the $N=4\times 4=16$ data significantly deviate from the larger system
results with $N=5\times 5\mp 1$. We suggest that the deviations from
scaling in $\epsilon_2(M)$ [Fig. \ref{fig2}(b)] and $\delta_2(M)$ [Fig. 
\ref{fig3}] arise
from peculiarities of the $4\times 4$ system \cite{fabricius91b} and that
the thermodynamical limit is fairly
well approximated by the larger system results.

%%%%%%%%%%%%%%%%BEGIN-FIGURE%%%%%%%%%%%%%%%%%
%\begin{figure}[ht]
%%%%%%%%%%%%%%%%%END-FIGURE%%%%%%%%%%%%%%%%%%

%%%%%%%%%%%%%%%%BEGIN-FIGURE%%%%%%%%%%%%%%%%%
%%%%%%%%%%%%%%%%%END-FIGURE%%%%%%%%%%%%%%%%%%

In the collinear antiferromagnetic phase $\alpha>\alpha_0(M)$ perturbation
theory in $\beta=1/\alpha$ to second order yields for the lowest eigenvalue
of $H_1+\alpha H_2$ in the sector with magnetization $M=s/N$:

\begin{eqnarray}
E(s,{\bf p}_s,\alpha) & = & N\left( 2M^2+\alpha\epsilon_1(M)+\alpha^{-1}
\delta_{-1}(M)\right)\,,\nonumber\\
 & & \quad\quad\quad\quad\quad\quad \alpha>\alpha_0(M)\,.\label{2nd_gt}
\end{eqnarray}

The perturbative results to first order -- i.e. the first two terms on
the right-hand side of \eqref{2nd_gt} -- are derived in App. A. Note that
this result can be  expressed in terms of the ground state energy per
site $\epsilon_1(M)$ of the nearest-neighbour model $H=H_1$! The second
order contribution $\delta_{-1}(M)$ is computed in App. B.

We would like to stress at this point, that the perturbation
expansions \eqref{2nd_lt} and \eqref{2nd_gt} hold in the 
thermodynamical limit.

The determination of the coefficients $\epsilon_1(M)$,
$\epsilon_2(M)$, $\delta_2(M)$ and $\delta_{-1}(M)$ on
finite clusters might be affected by finite-size effects;
their magnitude is illustrated in Figs. \ref{fig2}, \ref{fig3} 
and \ref{fig10} and can
be reduced by a computation on larger clusters and by a systematic
finite-size analysis. The coefficients $\epsilon_1(M)$ and
$\epsilon_2(M)$ are related to spin-spin correlators over
nearest and next-nearest neighbour sites in the unfrustrated
system ($\alpha=0$). Here, the finite-size effects are
small and well under control.

\begin{figure}[]
\centerline{\hspace{-3mm}\epsfig{file=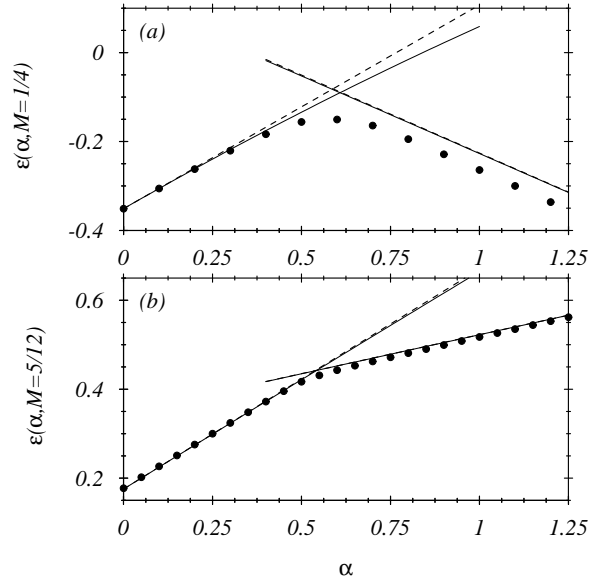,width=7.6cm,angle=0}}
\caption{Comparison of the $\alpha$-dependence of the ground state energy 
per site $\epsilon(\alpha,M)$\, -- $M=1/4$\,(a), $5/12$\,(b) --
and the perturbation theory results \eqref{2nd_lt}, \eqref{2nd_gt}.
The numerical data for $M=1/4$ was taken from a $4\times 4$ system
and for $M=5/12$ from a $6\times 6$ system, both with periodic boundary
conditions.}
\label{fig4}
\end{figure}

In Fig.~\ref{fig4}(a) we show the $\alpha$-dependence of 
ground state energies per
site $\epsilon(\alpha,M=1/4)$ as they follow from the perturbation
expansions \eqref{2nd_lt} and \eqref{2nd_gt}. The dashed lines 
represent the contributions to first order, which is fixed by the
coefficients
\begin{eqnarray}
\epsilon_1(M=1/4)=-0.34 & \,,\, & \epsilon_2(M=1/4)=0.46\nonumber\\
\delta_2=\delta_{-1}=0\,.\hspace{0.55cm} &  & \nonumber
\end{eqnarray}
These values can be read off Figs. \ref{fig2}(a), (b). The solid lines
include second order contributions. Indeed, second order contributions
seem to be relevant only in the antiferromagnetic phase \eqref{2nd_lt}.
They are negligible in the collinear antiferromagnetic phase \eqref{2nd_gt}.
The solid dots represent numerical results for a $4\times 4$ system.
Comparison with the perturbative results reveal small finite-size effects
in the antiferromagnetic phase \eqref{2nd_lt}, however large finite-size
effects in the collinear antiferromagnetic phase \eqref{2nd_gt}.

%%%%%%%%%%%%%%%%BEGIN-FIGURE%%%%%%%%%%%%%%%%%
%%%%%%%%%%%%%%%%%END-FIGURE%%%%%%%%%%%%%%%%%%

The comparison of the perturbation expansions \eqref{2nd_lt} and \eqref{2nd_gt}
with numerical results for $M=5/12$ on a larger system with $6\times 6=36$
sites is shown in Fig.~\ref{fig4}(b). Note in particular that the deviations in the
collinear antiferromagnetic phase are smaller here, owing to the larger
system size. In this case, the coefficients turn out to be
\begin{eqnarray}
\epsilon_1(M=5/12)=\,0.17 & \,,\, & \epsilon_2(M=5/12)=0.49\nonumber
\end{eqnarray}

The first order perturbation expansion \eqref{2nd_gt} in the collinear
antiferromagnetic phase $\alpha>\alpha_0(M)$ predicts the vanishing of
the $\alpha$-dependence for $M_0\simeq 0.37$, where $\epsilon_1(M_0\simeq
0.37)=0$ [cf. Fig.~\ref{fig2}(a)]. 

\begin{eqnarray}
\epsilon(M_0,\alpha) & = & 2M_0^2\quad\mbox{for }\alpha>\alpha_0(M_0)\,.
\end{eqnarray}

\begin{figure}[]
\centerline{\hspace{-3mm}\epsfig{file=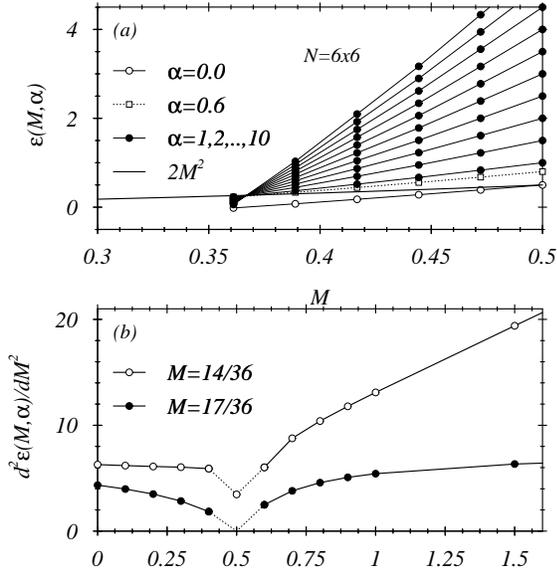,width=7.6cm,angle=0}}
\caption{Ground state energy per site -- $\epsilon(M,\alpha)$ vs. $M$ -- (a) and
its second derivative -- $\partial^2\epsilon/\partial M^2$ vs. $\alpha$ -- (b).}
\label{fig5}
\end{figure}

Fig.~\ref{fig5}(a) shows the $M$-dependence of $\epsilon(M,\alpha)$ on a
$6\times 6$ system for various values of $\alpha$. All curves with
$\alpha>0.6$ -- where the system is in the collinear antiferromagnetic
phase -- coincide at $M_0\simeq 0.37$. Data points for $\alpha=0$ (open
circles) -- i.e. in the antiferromagnetic phase -- do not share this
property!

The second derivative of $\epsilon(M,\alpha)$ with respect to $M$,
computed from finite system results with $N=36$, $\Delta M=1/N$:

\begin{eqnarray}
\epsilon^{(2)}=\frac{\partial^2\epsilon}{\partial M^2} & = &
\frac{1}{(\Delta M)^2}\Big(\epsilon(M+\Delta M,\alpha)+\nonumber\\
 & & \quad +\epsilon(M-\Delta M,\alpha)-2\epsilon(M,\alpha)\Big)
\end{eqnarray}

is shown in Fig.~\ref{fig5}(b).

The data points with $M=1/2-1/N$, $N=36$ (one spin flipped) have
a dip at $\alpha=1/2$ indicating that the second derivative might
vanish here in the thermodynamical limit $N\rightarrow\infty$.
Indeed, we see from a Taylor expansion around $M=M_0$:

\begin{eqnarray}
\epsilon(M,\alpha) & = & \epsilon(M_0,\alpha)+\epsilon^{(1)}(M-M_0)
+\frac{1}{2!}\epsilon^{(2)}(M-M_0)^2\nonumber\\
 & & +\frac{1}{3!}\epsilon^{(3)}
(M-M_0)^3+\ldots ,\label{taylor}\\
\epsilon^{(k)} & = & \left.\frac{\partial^k\epsilon}{\partial M^k}
\right|_{M=M_0}\label{taylor2}
\end{eqnarray}

that the vanishing of the second derivative $\epsilon^{(2)}=0$
induces a square root singularity in the magnetization curve

\begin{eqnarray}
|M-M_0|\cdot \left|\frac{\epsilon^{(3)}}{2}\right| & = &
|B-B_0|^{1/2}\,,\label{square}
\end{eqnarray}
provided that the third derivative $\epsilon^{(3)}$ does not vanish.

%%%%%%%%%%%%%%%%BEGIN-FIGURE%%%%%%%%%%%%%%%%%
%%%%%%%%%%%%%%%%%END-FIGURE%%%%%%%%%%%%%%%%%%

%Such a singular behaviour has been observed for $\alpha=1/2$ and
%$M_0=1/2$ [Ref. \onlinecite{yang97}] on system sizes $5\times 5\pm 1=26,\,24$,
%$7\times 7\pm 1=50,\,48$ with helical boundary conditions.

Under this assumption the numerical results for the magnetization
curve at $\alpha=1/2$ and $M\rightarrow 1/2$ on system sizes
$5\times 5\pm 1=26,24$, $7\times 7\pm 1=50,48$ with helical
boundary conditions could be described in Ref. \onlinecite{yang97}
by the square root \eqref{square}.

However, it has been pointed out by Honecker \cite{honecker00b} 
in a recent publication,
that numerical results on systems of size $6\times 6$, $8\times 8$ 
with periodic
boundary conditions indicate that the magnetic fields near saturation
\begin{eqnarray}
B(M=1/2(1-n/N),\alpha=1/2,N) & = & 4\quad\nonumber\\
 & & \label{peculiar} \\[-5pt]
\mbox{for}\quad n=0,1,2,..L,\quad N=L^2 & & \nonumber
\end{eqnarray}
are independent of $n$. This means that the system jumps discontinuously
from the magnetization $M=1/2\cdot(1-L/N)$ into the fully magnetized state 
$M=1/2$. The size of the jump $\Delta M=L/N$ vanishes in the thermodynamical
limit, so one expects a smooth approach to saturation in this limit.
%Whether or not this approach is of the form \eqref{square} remains an
%open question.
Moreover the behaviour \eqref{peculiar} suggests, that all the
derivatives \eqref{taylor2} near saturation $M_s=1/2$ vanish,
signalling an essential singularity.

%This would mean that the system jumps
%discontiniously into the fully magnetized state ($M=M_s=1/2$) -- in
%contrast to the square root behaviour predicted by \eqref{square}.

Going to a smaller value of $M$ ($M=14/36=0.39$) we find again a
dip in the second derivative. However, it is very difficult to
locate the dip position with our finite system results. 
%Indeed we
%suggest that the dip position coincides with the phase boundary
%$\alpha_0(M)$, which meets the point $\alpha_0(M=1/2)=1/2$ at
%saturaion $M=1/2$.

%%%%%%%%%%%%%%%%%%%%%%%%%%%%%%%%%%%%%%%%%%%%%%%%%%%%%%%%%%%%%%%%%%%%%%%%

\section{Free energies and magnetization curves}

According to eq. \eqref{b_m} the magnetization curves follow from a
minimum of the ``generalized free energy'' per site
\begin{eqnarray}\label{free_energy}
f(M,\alpha) & = & \epsilon(M,\alpha)-M\cdot B
\end{eqnarray}
with respect to $M$ at fixed $B$. The perturbation expansions
\eqref{2nd_lt} and \eqref{2nd_gt} therefore yield for the magnetization
curves in the antiferromagnetic and collinear antiferromagnetic phase:

\begin{eqnarray}
B(M,\alpha) & = & \frac{d\epsilon_1}{dM}+\alpha\frac{d\epsilon_2}{dM}
+\alpha^2\frac{d\delta_2}{dM}
\label{bm_lt}\\[5pt]
B(M,\alpha) & = & 4M+\alpha\frac{d\epsilon_1}{dM}
+\alpha^{-1}\frac{d\delta_{-1}}{dM}
\label{bm_gt}\,.
\end{eqnarray}

\begin{figure}[]
\centerline{\hspace{6mm}\epsfig{file=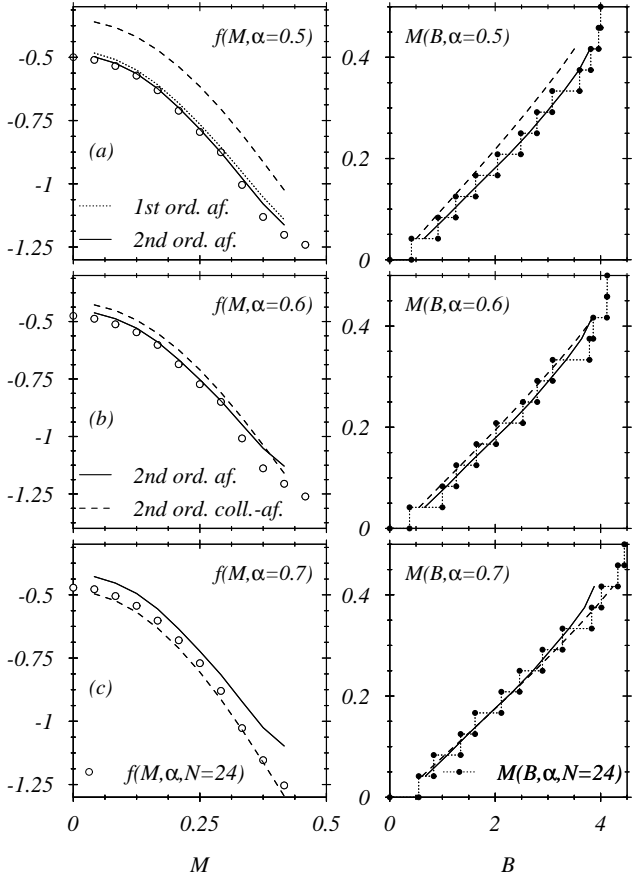,width=8.0cm,angle=0}}
\caption{Free energies and magnetization profiles for $\alpha=0.5(a)$,
$0.6(b)$ and $0.7(c)$. Solid and dashed curves show the second order
perturbation results [cf. \eqref{bm_lt} and \eqref{bm_gt}] in the 
antiferromagnetic
and collinear antiferromagnetic regime, respectively. The circles
represent exact diagonalization results on a $N=5\times 5-1=24$ lattice
with helical boundary conditions.}
\label{fig6}
\end{figure}

In Fig.~\ref{fig6} we present on the left-hand side the free 
energies \eqref{free_energy}
-- as they follow from \eqref{2nd_lt}, \eqref{2nd_gt} -- for $\alpha=0.5,\,0.6,
\,0.7$.

For $\alpha=0.5$ [Fig.~\ref{fig6}(a)] we are definitely in the 
antiferromagnetic phase, since the corresponding free energy here
(solid line) is below the one in the collinear antiferromagnetic
phase (dashed line). In order to demonstrate the quality of the
perturbation expansion \eqref{2nd_lt}, we also plotted the first
order contribution alone (dotted line) and the numerical data
(open circles) on a finite system with $N=5\times 5-1=24$ sites
(with helical boundary conditions). Good agreement is found for
$M\leq 0.3$. For larger $M$-values, the numerical data are below
the perturbative results.

The corresponding magnetization curves at $\alpha=0.5$ can be 
seen on the right-hand side. Again, the first perturbation expansion
\eqref{bm_lt} (solid line) follows the numerical results (step 
function), whereas the second perturbation expansion \eqref{bm_gt} 
(dashed curve) lies above the numerical results.

%%%%%%%%%%%%%%%%BEGIN-FIGURE%%%%%%%%%%%%%%%%%
%%%%%%%%%%%%%%%%%END-FIGURE%%%%%%%%%%%%%%%%%%

For $\alpha=0.7$ [Fig.~\ref{fig6}(c)] we are definitely in the
collinear antiferromagnetic phase. The corresponding free energy
(dashed line) is now below the one of the antiferromagnetic phase
(solid line). The numerical data are close to the dashed line.
In the magnetization curve (right-hand side) the perturbation 
expansions \eqref{bm_lt}, \eqref{bm_gt} and the numerical results
(step function) coincide for $M<1/3$.

In between ($0.5<\alpha<0.7$) we therefore expect the phase
transition. In Fig.~\ref{fig6}(b) we show the situation for
$\alpha=0.6$. For $M\leq 1/3$ and $M>1/3$, the free energies are
minimal in the antiferromagnetic and collinear antiferromagnetic
phase, respectively. The numerical data (open circles) on the
finite system with $N=24$ sites are lying below the
perturbative results. The largest deviations are found in the
collinear antiferromagnetic phase ($M>1/3$) indicating that 
both perturbation expansions and finite system results become
questionable.

%%%%%%%%%%%%%%%%BEGIN-FIGURE%%%%%%%%%%%%%%%%%
\begin{figure}[ht!]
\centerline{\hspace{-3mm}\epsfig{file=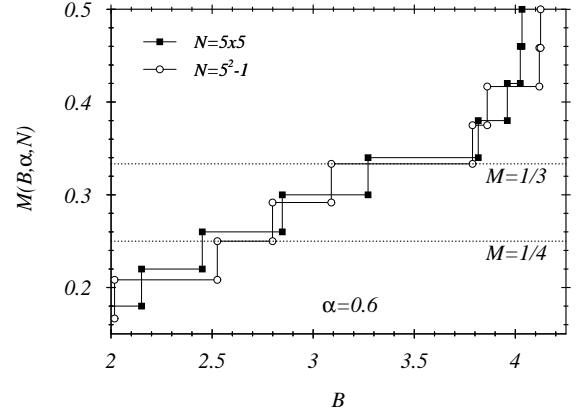,width=8.0cm,angle=0}}
\caption{Behaviour of the magnetization curves of systems with
periodic ($N=5\times 5$) and helical ($N=5\times 5-1$)
boundary conditions around magnetization $M=1/4,1/3$ and frustration
$\alpha=0.6$}
\label{fig6c}
\end{figure}
%%%%%%%%%%%%%%%%%END-FIGURE%%%%%%%%%%%%%%%%%%

For small values of $M$ ($M<1/3$) the numerical data for the
magnetization curve (right-hand side) are well 
reproduced by the perturbation expansion \eqref{bm_lt} for the
antiferromagnetic phase. 
At $M=1/3$ a plateau-like structure seems to evolve in the
in the numerical data for $N=5\times 5-1=24$ with helical
boundary conditions. Unfortunately, we cannot check whether
this structure will survive on larger systems with helical
boundary conditions like $N=7\times 7-1=48$. However, the
structure seems to appear as well (at $M=0.34$) on a square 
lattice $N=5\times 5=25$ with periodic boundary conditions
as is demonstrated in Fig. \ref{fig6c}.

%At $M=1/3$ a plateau seems to evolve
%(in the numerical data) which of course cannot be described
%by the perturbation expansions \eqref{bm_lt} or \eqref{bm_gt}.
%Formation of gaps and plateaus are definitely non-perturbative
%effects.

%It is remarkable to note that the numerical results for the
%magnetization curve seem to indicate a singular behaviour, if
%we approach the (saturation) plateau at $M=1/2$ from below and
%the plateau at $M=1/3$ from above. A square root singularity
%of the type \eqref{2nd_gt} has been found for $M=1/2$. \cite{yang97}
%We suggest that this type of singularity dominates as well the
%behaviour of the magnetization curve near the upper critical
%field at $M=1/3$.

%The magnetization curve for $M\geq 1/4$ and $\alpha=0.6$ has been
%investigated in \cite{zhitomirsky00} for systems $4\times 4$, $4\times 6$
%and $6\times 6$ and periodic boundary conditions. A plateau has
%been found for $M=1/4$, which we do not see in our data for
%$N=5\times 5-1=24$ with helical boundary conditions. On the other hand,
%the plateau at $M=1/3$ in our data seems to be absent on the $6\times 6$
%system (with periodic boundary conditions).

%%%%%%%%%%%%%%%%%%%%%%%%%%%%%%%%%%%%%%%%%%%%%%%%%%%%%%%%%%%%%%%%%%%%%%%%

\section{Dynamical structure factors and frequency moment sum rules}
\label{}

In this section we will study the impact of the transition from
antiferromagnetic to collinear antiferromagnetic order on the
dynamical structure factors:

\begin{eqnarray}
S_{jj}({\bf q},\omega,M,\alpha) & = & \frac{1}{N}\sum_n\delta(E_n-E_0-
\omega)|\langle n|S_j({\bf q})|0\rangle|^2\nonumber\\
 & & \label{s_jj}
\end{eqnarray}

and their frequency moments:

\begin{eqnarray}
K^{(j)}_n({\bf q},M,\alpha) & = & \int_0^{\infty}\,d\omega\,\omega^n 
S_{jj}({\bf q},\omega,M,\alpha)\,.\label{k_j}
\end{eqnarray}

In particular, we expect signatures in the $\alpha$-dependence of the
structure factors at momenta ${\bf q}=(\pi,\pi)$ and ${\bf q}=(\pi,0)$.

\begin{figure}[]
\centerline{\hspace{-3mm}\epsfig{file=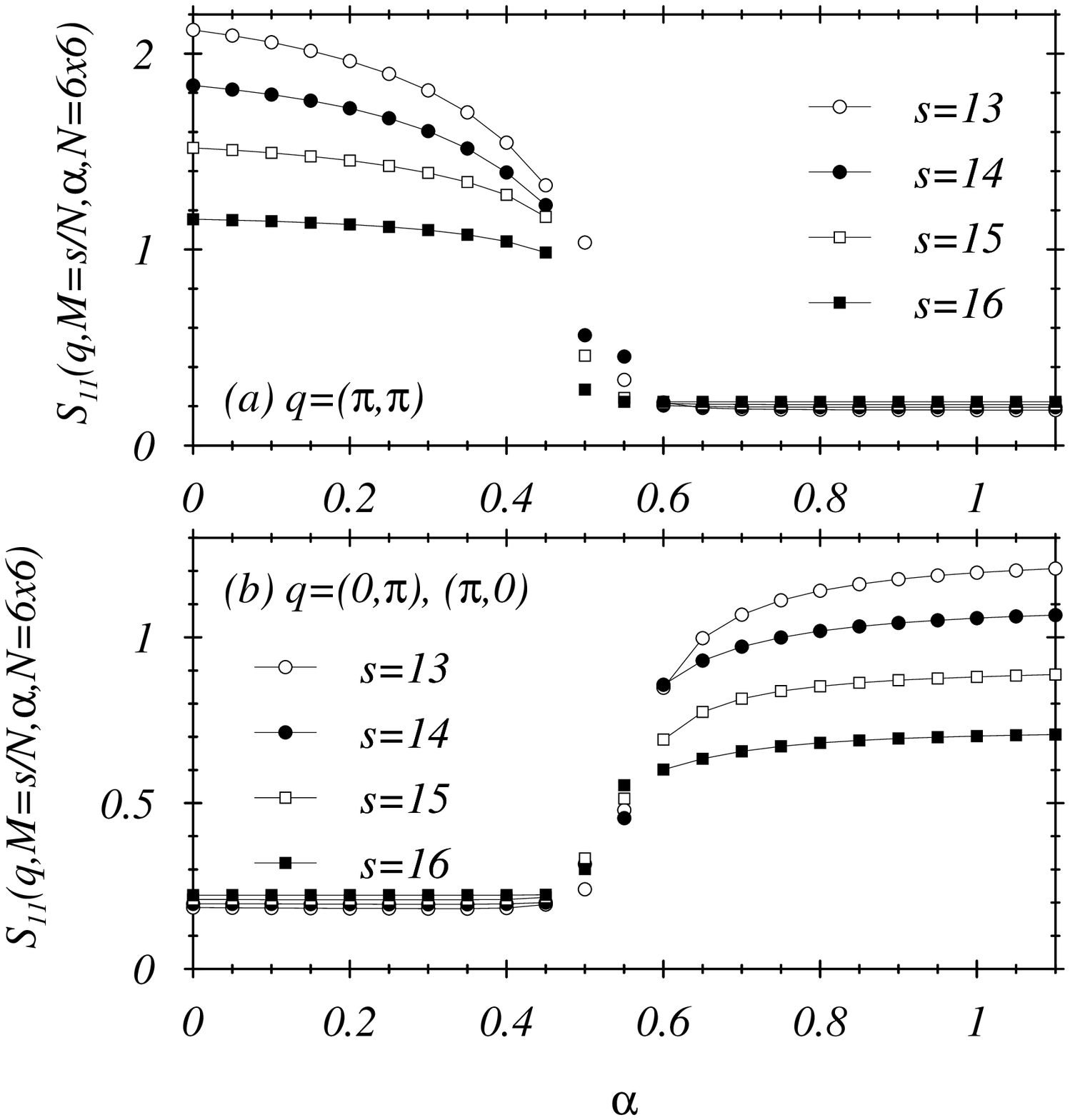,width=7.6cm,angle=0}}
\caption{Zeroeth transverse frequency moment 
(transverse static structure factor)
$K^{(1)}_0({\bf q},M,\alpha)=S_{11}({\bf q},M,\alpha)$ for system
size $N=6\times 6$ for different magnetizations and momenta
${\bf q}=(\pi,\pi)$(a) and ${\bf q}=(0,\pi),(\pi,0)$(b).}
\label{fig7}
\end{figure}

\begin{figure}[]
\centerline{\hspace{-3mm}\epsfig{file=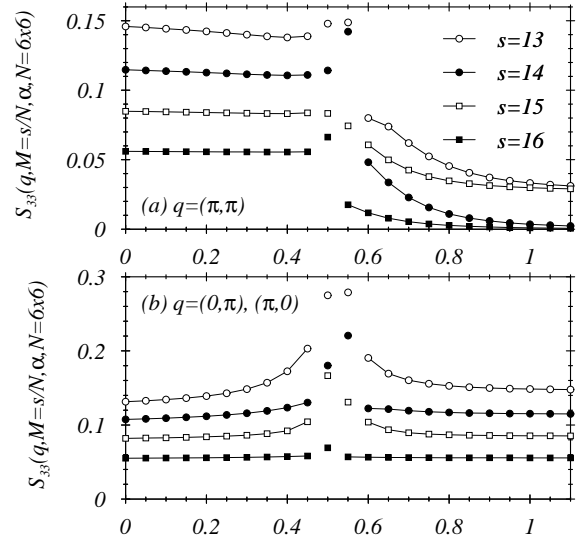,width=7.6cm,angle=0}}
\caption{Zeroeth longitudinal frequency moment 
(longitudinal static structure factor)
$K^{(3)}_0({\bf q},M,\alpha)=S_{33}({\bf q},M,\alpha)$ for system
size $N=6\times 6$ for different magnetizations and momenta
${\bf q}=(\pi,\pi)$(a) and ${\bf q}=(0,\pi),(\pi,0)$(b).}
\label{fig8}
\end{figure}

In Figs.~\ref{fig7} and \ref{fig8} we show the $\alpha$-dependence 
of the transverse and longitudinal static structure factors on a square
lattice with $N=6\times 6=36$ sites:

\begin{eqnarray}
K_0^{(j)}({\bf q},M,\alpha,N=36) & & \quad\mbox{for } 
M=\frac{1}{36}\cdot\Big(13,14,15,16\Big)\,.\label{k0_j}
\nonumber\\
\end{eqnarray}

%%%%%%%%%%%%%%%%BEGIN-FIGURE%%%%%%%%%%%%%%%%%
%%%%%%%%%%%%%%%%%END-FIGURE%%%%%%%%%%%%%%%%%%

The transverse structure factors ($j=1)$ for the momenta ${\bf q}=
(\pi,\pi)$ [Fig.~\ref{fig7}(a)] and ${\bf q}=(\pi,0)$, $(0,\pi)$
[Fig.~\ref{fig7}(b)] reveal a striking similarity in their
$\alpha$-dependence: All curves for ${\bf q}=(\pi,\pi)$ and
${\bf q}=(\pi,0)$, $(0,\pi)$ show a sensitive change in their
$M$- and $\alpha$-dependence when entering
the regimes $\alpha>0.6$ and $\alpha<0.5$, respectively.
%In the complementary regimes this sensitivity almost vanishes.

%are almost $M$-independent in the
%antiferromagnetic ($\alpha<0.5$) and collinear antiferromagnetic
%($\alpha>0.6$) phase, respectively. In contrast, a sensitive
%$M$-dependence appears for ${\bf q}=(\pi,\pi)$ and 
%${\bf q}=(\pi,0)$, $(0,\pi)$ in the complementary regimes
%$\alpha>0.6$ and $\alpha<0.5$, respectively.

%A sharp drop is visible in the transverse structure factor with
%${\bf q}=(\pi,\pi)$, which approaches a constant value in the
%collinear antiferromagnetic phase:

In particular we see in our data, that

\begin{eqnarray}
S_{11}({\bf q}=(\pi,\pi),M,\alpha) & \simeq\quad M/2  & 
\quad\mbox{for }
\alpha>0.6\,.
\end{eqnarray}

This behaviour is a consequence of the ground state wave function
\eqref{product}.

The longitudinal structure factor [\eqref{k0_j} for $j=3$] at the
momenta ${\bf q}=(\pi,\pi)$ [Fig.~\ref{fig8}(a)] and
${\bf q}=(\pi,0)$, $(0,\pi)$ [Fig.~\ref{fig8}(b)] look similar
in the small-$\alpha$ regime ($\alpha<0.5$): One observes a
constant behaviour with $\alpha$ and a monotonic decrease with $M$.
For large values $\alpha$, the longitudinal structure factors
behave quite differently: For ${\bf q}=(\pi,\pi)$ the curves with
$M=s/N$, $s$ even and $s$ odd approach different limiting values,
respectively. This feature is not visible for ${\bf q}=(\pi,0)$,
$(0,\pi)$.

Numerical data for the static structure factor on a $6\times 6$ system
with periodic boundary conditions have been presented in. \cite{zhitomirsky00}
At fixed magnetization $M=1/4$ -- where the authors see a plateau in
the magnetization curve -- the data points for the longitudinal
structure factor at ${\bf q}=(\pi,\pi)$ and ${\bf q}=(\pi,0)$ show
a broad maximum in the regime $0.55\leq\alpha\leq 0.67$, which is
also visible in our results for larger $M$ values 
[cf. Figs. \ref{fig8}(a), (b)
in the interval $0.5<\alpha <0.6$].

%%%%%%%%%%%%%%%%BEGIN-FIGURE%%%%%%%%%%%%%%%%%
%%%%%%%%%%%%%%%%%END-FIGURE%%%%%%%%%%%%%%%%%%

Let us next turn to the first frequency moment. As was pointed out
by Hohenberg and Brinkman \cite{hohenberg75} and M\"uller
\cite{mueller82} it can be expressed in terms of ground state expectation
values for two-fold commutators:

\begin{eqnarray}
K^{(j)}_1({\bf q},M,\alpha) & = & \frac{1}{N}\sum_n\left(E_n-E_0-B(S_{3n}-
S_{30})\right)\nonumber\\
 & & \times |\langle n|S_j({\bf q})|0\rangle|^2\nonumber\\
 & = & \frac{1}{N}\langle 0|S^+_j({\bf q})HS_j({\bf q})-HS^+_j({\bf q})
S_j({\bf q})|0\rangle\nonumber\\
 & = & \frac{1}{2N}\langle 0|[[S^+_j({\bf q}),H],S_j({\bf q})]|0\rangle\,.
\end{eqnarray}

$S_{30}$ and $S_{3n}$ are the 3-component of the total spin in the ground
state $|0>$ and the excited states $|n>$, respectively.
The calculation of the commutators yields:

\begin{eqnarray}\label{mom_1}
K^{(j)}_1({\bf q},M,\alpha) & = & -(1-\cos q_1)C_j(\hat{1})-
(1-\cos q_2)C_j(\hat{2})\nonumber\\
 & & -\alpha\left(1-\cos(q_1+q_2)\right)C_j(\hat{1}+\hat{2})\nonumber\\
 & & -\alpha\left(1-\cos(q_1-q_2)\right)C_j(\hat{1}-\hat{2})\nonumber\\
 & & -\frac{1}{2}M_jB(M,\alpha)\,,
\end{eqnarray}

where

\begin{eqnarray}
C_j(\hat{\nu}) & = & \frac{1}{N}\sum_{{\bf x}}\left(\langle 0|{\bf S}({\bf x}){\bf S}
({\bf x}+\hat{\nu})|0\rangle\right.\nonumber\\
 & & \quad\quad\quad\left.-\langle 0|S_j({\bf x})S_j({\bf x}+\hat{\nu})|0\rangle\right)
\end{eqnarray}

for $j=1,2,3$, $\hat{\nu}=\hat{1}$, $\hat{2}$, $\hat{1}+\hat{2}$,
$\hat{1}-\hat{2}$ and

\begin{center}
\begin{tabular}{lll}
$M_j=M$ & for & $j=1,2$\,,\\
$M_j=0$ & for & $j=3$\,.
\end{tabular}
\end{center}

It should be noted that the dependence on the momentum transfer
${\bf q}=(q_1,q_2)$ is completely given by the Fourier factors on the
right-hand side of \eqref{mom_1}. The physical meaning of the first
frequency moment can be seen from the definitions \eqref{s_jj} and
\eqref{k_j}.

%%%%%%%%%%%%%%%%BEGIN-FIGURE%%%%%%%%%%%%%%%%%
\begin{figure}[]
\centerline{\hspace{-3mm}\epsfig{file=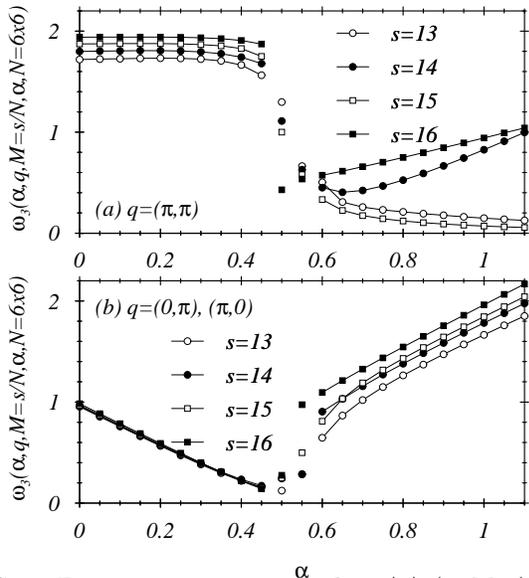,width=7.6cm,angle=0}}
\caption{Frequency expectation values $\langle \omega\rangle_3({\bf q},M,\alpha)$
for a $N=6\times 6$ spin system for different magnetizations and
${\bf q}=(\pi,\pi)$(a) and ${\bf q}=(0,\pi),(\pi,0)$(b).}
\label{fig9}
\end{figure}
%%%%%%%%%%%%%%%%%END-FIGURE%%%%%%%%%%%%%%%%%%

The ratio

\begin{eqnarray}\label{omega_j}
< \omega_j>({\bf q},M,\alpha) & = & \frac{K^{(j)}_1({\bf q},M,\alpha)}
{K^{(j)}_0({\bf q},M,\alpha)}\,,\quad\quad j=1,3
\end{eqnarray}

yields the average excitation energy in the spectrum of states which can
be reached from the ground state $|0\rangle =|{\bf p}_s,M_s\rangle$ by means of the
operator $S_j({\bf q})$ $j=1,2,3$. This ratio is shown for the
longitudinal case $j=3$ and magnetizations $M=1/36\cdot(13,14,15,16)$
on a $6\times 6$ system in Fig.~\ref{fig9}.

At momentum ${\bf q}=(\pi,\pi)$ [Fig.~\ref{fig9}(a)], we observe
a sharp drop in the width of the excitation spectrum at $\alpha\approx
0.5$, i.e. close to the phase boundary.

At ${\bf q}=(\pi,0)$, $(0,\pi)$ [Fig.~\ref{fig9}(b)] the longitudinal
structure factors for all $M$-values follow the same straight line, which
ends at a minimum at $\alpha=0.5$.

%***************************************************************************

\section{Conclusions and discussion}

The magnetization process in the two-dimensional $J_1-J_2$ model is
well described in the antiferromagnetic phase ($\alpha<0.5$) by a
perturbation expansion around $\alpha=0$ and in the collinear
antiferromagnetic phase ($\alpha>0.7$) by a perturbation expansion
around $\alpha^{-1}=0$. In these regimes, the magnetization
curves are smooth, monotonically increasing and convex. The change in the
magnetic order can be seen in the $\alpha$-dependence of the transverse 
and longitudinal structure factors at momenta ${\bf q}=(\pi,\pi)$ and
${\bf q}=(\pi,0),(0,\pi)$, respectively.

Using frequency moment sum rules we have also studied the variation of
the average excitation energy $\langle \omega\rangle$ \eqref{omega_j} with 
$\alpha$. Approaching
the transition region from antiferromagnetic to collinear antiferromagnetic
order, the spectral weight is more and more concentrated around low
frequencies.

Between the antiferromagnetic and collinear antiferromagnetic phase
-- i.e. for $0.5<\alpha<0.7$ -- a phase of transverse disorder has been
suggested. \cite{zhitomirsky00} 

%Here one expects plateaus and perhaps other characteristic features
%in the magnetization curve.
%
%Unfortunately, numerical results obtained on finite clusters
%$N=4\times 4=16$, $4\times 6=24$, $6\times 6=36$, $8\times 8=64$
%with periodic boundary conditions \cite{zhitomirsky00,honecker00b}
%and $N=5\times 5\pm 1=24,26$, $7\times 7\pm 1=48,50$ with helical
%boundary conditions \cite{yang97} lead to different signatures
%on the behaviour of the magnetization curve in the thermodynamic
%limit.

This regime is not accessible with the perturbative methods
developed in this paper; nonperturbative effects -- like
plateaus -- are expected here.

To study these, we have to rely on numerical results on finite
clusters. The results available so far reveal a strong dependence
on the size of the clusters and the boundary conditions. For the
moment we meet the following situation:

(A) At $\alpha=1/2$ the approach to saturation ($M_s=1/2$) is no 
longer linear but appears to develop a singularity. The type of
the singularity is not yet clear:

On one hand, numerical data on square lattices $6\times 6=36$,
$8\times 8=64$ with periodic boundary conditions reveal the
peculiar property \eqref{peculiar} for the magnetic fields
near saturation. If we assume in the thermodynamic limit a
series expansion [cf. \eqref{taylor}] of the energies per site
$\epsilon(M,\alpha)$ at the saturation point $M_s=1/2$,
infinitely many derivatives [cf. \eqref{taylor2}] would vanish,
indicating an essential singularity.

%On the other hand, our numerical results obtained on lattices
%$5\times 5\pm 1$, $7\times 7\pm 1$ with helical boundary
%conditions only indicate that the second term $\epsilon^{(2)}
%(M_s,\alpha)$ in the Taylor expansion vanishes for $\alpha=1/2$.
%Provided that the third order term $\epsilon^{(3)}(M_s,\alpha)$
%does not vanish, the magnetization curve develops a square
%root singularity.

On the other hand, numerical results [Ref. \onlinecite{yang97}] on finite
lattices $5\times 5\pm 1=26,24$, $7\times 7\pm 1=50,48$ with helical
boundary conditions do not share the property \eqref{peculiar} for
the magnetic fields near saturatiion $M\rightarrow 1/2$, $\alpha=1/2$.
It was suggested in Ref. \onlinecite{yang97} that the magnetization
curve develops a square root singularity, which follows if the
second derivative $\epsilon^{(2)}(M,\alpha=1/2)$ in \eqref{taylor}
vanishes for $M\rightarrow 1/2$, but not the higher ones
$\epsilon^{(k)}(M,\alpha=1/2)$ $k=3,4,\ldots$.

(B) At $\alpha=0.6$ strong evidence for a plateau at $M=1/4$ has been
reported by Honecker et al. \cite{zhitomirsky00,honecker00b}
who performed a finite-size analysis of exact diagonalization
results on 3 clusters $4\times 4$, $4\times 6$ and $6\times 6$
with periodic boundary conditions. They do not find any signature
for a plateau at $M=1/3$. 
In our results obtained from a $N=5\times 5-1=24$ system with
helical boundary conditions we do not find a plateau at $M=1/4$;
but there are indications for a plateau-like structure at $M=1/3$.
We are aware of the fact, that this discrepancy might come
from special finite-size effects due to the helical boundary
conditions, which only disappear for much larger systems (e.g.
$N=7\times 7-1=48$, ..). However, these are not yet accessible.
%for the moment.

In order to test the dependence on the boundary conditions we
have compared the magnetization curves for $N=5\times 5-1=24$
with helical boundary conditions and for $N=5\times 5=25$
with periodic boundary conditions. The results (Fig. \ref{fig6c})
look similar. This would mean that the plateau structures in
the magnetization curve has a sensitive dependence on the
system size $N=L\times L$; in particular results for $L$ even
and $L$ odd might be quite different.
Therefore, the precise determination of the positions and widths
of the plateaus in the frustrated Heisenberg model is indeed
a very delicate problem.

%The sensitive dependence on the boundary conditions and system size
%demonstrates, that the precise determination of the plateau and
%the widths of the plateaus in the frustrated Heisenberg model is 
%indeed a very delicate problem. 
A better theoretical understanding
of the magnetic order and of the mechanisms which create the
plateaus is really needed.

%***************************************************************************

\acknowledgements

We would like to thank M. Karbach for discussion and a critical
reading of the manuscript.

%***************************************************************************

\begin{appendix}
\section{The strong frustration limit}\label{strong_al}
As was pointed out in Sec. I [cf. \eqref{h+-}], the diagonal couplings
build up two independent nearest-neighbour Hamiltonians on the even and odd 
sublattice, respectively. 

We start with the eigenvalue equations for these sublattices

\begin{eqnarray}\label{states}
\left.\begin{array}{c}H_1^{(\pm)}\\ {\bf S}^2\\ S_3\end{array}\right\}
\Psi^{(\pm)}(\sigma,{\bf p}_{\sigma}) & = & \left.\begin{array}{c}
E(\sigma,{\bf p}_{\sigma})\\ \sigma(\sigma+1)\\ \sigma\end{array}
\right\} \Psi^{(\pm)}(\sigma,{\bf p}_{\sigma})\,.
\end{eqnarray}

In particular we are interested in the ground state with definite
(squared) total spin and total magnetization $M=\sigma/(N/2)$. In the
thermodynamical limit $N/2\rightarrow\infty$ the ground state momenta
${\bf p}_{\sigma}$ follow Marshall's sign rule \cite{marshall55}.
If $N/4+\sigma$ is even, the ground state momentum turns out to be
${\bf p}_{\sigma}=(0,0)$; if $N/4+\sigma$ is odd, one obtains
${\bf p}_{\sigma}=(\pi,\pi)$. The lowest eigenstates of the 
Hamiltonian $H_2=H_1^{(+)}+H_1^{(-)}$ with even total spin and zero
total momentum ${\bf p}=(0,0)$ can be constructed as a product of states
of the lowest eigenstates \eqref{states} of the subsystems:

\begin{eqnarray}\label{product}
\Psi(s,{\bf p}_s=(0,0),N) & = & \Psi^{(+)}(s/2,{\bf p}_{s/2},N/2)
\nonumber\\
 & & \times\Psi^{(-)}(s/2,{\bf p}_{s/2},N/2)\,.
\end{eqnarray}

The two subsystems defined on the even and odd sites can be transformed
into each other via a translation of one lattice spacing. We can
therefore impose the following identification on the wave functions
with identical quantum numbers in \eqref{product}

\begin{eqnarray}\label{psi_m}
\Psi^{(-)}(\sigma,{\bf p}_{\sigma},N/2) & = & T_1
\Psi^{(+)}(\sigma,{\bf p}_{\sigma},N/2),\quad\sigma=1/2\,.
\end{eqnarray}

\eqref{psi_m} guarantees that the product states \eqref{product} are
invariant under both ($\nu=1,2$) translation operators:

\begin{eqnarray}\label{psi_t}
T_{\nu}\Psi(s,{\bf p}_s=(0,0),N) & = & 
\Psi(s,{\bf p}_s=(0,0),N)\,,
\end{eqnarray}

i.e. they are momentum zero eigenstates. For the proof of this 
statement one has to realize that the eigenstates on the subsystems
behave as follows under translations:

\begin{eqnarray}
T_1T_2\Psi^{(\pm)}(\sigma,{\bf p},N/2) & = & e^{ip_1}\cdot\Psi
^{(\pm)}(\sigma,{\bf p},N/2)\nonumber\\
T_1T_2^{-1}\Psi^{(\pm)}(\sigma,{\bf p},N/2) & = & e^{ip_2}\cdot\Psi
^{(\pm)}(\sigma,{\bf p},N/2)\label{a5}\\
T_1^2\Psi^{(\pm)}(\sigma,{\bf p},N/2) & = & e^{i(p_1+p_2)}\cdot\Psi
^{(\pm)}(\sigma,{\bf p},N/2)\nonumber
%T_{1}^2\Psi^{(\pm)}(\sigma,{\bf p},N/2) & = & 
%\Psi^{(\pm)}(\sigma,{\bf p},N/2)\label{a6}
\end{eqnarray}

It should be noted that the product states \eqref{product} where the total
spin $s$ is distributed to equal parts on the two sublattices indeed yields
the ground state with energy 

\begin{eqnarray}
E(s,{\bf p}_s=(0,0),N) & = & 2E(s/2,{\bf p}_{s/2},N/2)\,.
\end{eqnarray}

Any other distribution, e.g. with a total spin $s/2+\Delta s$ on the even
and $s/2-\Delta s$ on the odd lattice will lead to an eigenstate with
higher energy. This is a consequence of the fact that the lowest energy
is monotonically increasing and a convex function of the magnetization
$M=s/N$.

The construction of eigenstates of $H_1^{(+)}+H_1^{(-)}$ with odd
total spin $s$ and total momentum ${\bf p}_s=(\pi,0)$, $(0,\pi)$ is
more involved. In this case an equal distribution of the total spin
on the two sublattices is impossible since the sublattice spin $\sigma$
has to be integer. The state with lowest energy is found with the ansatz:

\begin{eqnarray}
\Psi(s,{\bf p}_s=(\pi,0),N) & = & \quad\quad\quad\quad\quad\quad\quad
\quad\quad\quad\quad\quad\nonumber
\end{eqnarray}
\vspace{-0.7cm}
\begin{eqnarray}\label{ansatz}
\frac{1}{\sqrt{2}}\left(\Psi^{(+)}
(\sigma,{\bf p}_{\sigma},N/2)\Psi^{(-)}(\sigma-1,{\bf p}_{\sigma-1},N/2)
\right. & & \nonumber\\
\quad\quad\quad -\left.\Psi^{(+)}(\sigma-1,{\bf p}_{\sigma-1},N/2)\Psi^{(-)}(\sigma,
{\bf p}_{\sigma},N/2)
\right) & & \label{a8}
\end{eqnarray}

where 

\begin{eqnarray}
\sigma & = & \frac{s+1}{2}\,.
\end{eqnarray}

Imposing the identification \eqref{psi_m} on the ground states with equal
quantum numbers, one finds the following transformation properties of the
states \eqref{ansatz}:

\begin{eqnarray}
T_1\Psi(s,{\bf p}_s=(\pi,0),N) & = & -\Psi(s,{\bf p}_s=(\pi,0),N)\nonumber\\
 & & \\[-6pt]
T_2\Psi(s,{\bf p}_s=(\pi,0),N) & = & \Psi(s,{\bf p}_s=(\pi,0),N)\nonumber
\end{eqnarray}

which means that the ground state momentum is ${\bf p}_s=(\pi,0)$. The
degenerate ground state with momentum ${\bf p}_s)=(0,\pi)$ is obtained
if we substitute the translation operator $T_1$ by $T_2$ in \eqref{psi_m}.

In the strong coupling limit $\alpha\rightarrow\infty$ ($\beta=1/\alpha
\rightarrow 0$), the nearest-neighbour operator $H_1$ in \eqref{h0} acts
as a perturbation operator and we therefore have to compare in first
order the expectation values of $H_1$ between the ground states 
\eqref{product} and \eqref{a8}, respectively:

\begin{eqnarray}
 & & \langle \Psi(s,{\bf p}_s=(0,0),N)|H_1|\Psi(s,{\bf p}_s=(0,0),N)\rangle\nonumber\\
 & = & \sum_{{\bf x}\in (+)}\langle \Psi^{(+)}(\sigma)|S_3({\bf x})|\Psi^{(+)}
(\sigma)\rangle
\nonumber\\
 & & \times\sum_{{\bf \nu}}\langle \Psi^{(-)}(\sigma)|T_{{\bf \nu}}S_3({\bf x})
T_{{\bf \nu}}^{-1}
|\Psi^{(-)}(\sigma)\rangle\,.
\end{eqnarray}

Here we have used the fact that the nearest-neighbour couplings in
$H_1$ connect the spin operator ${\bf S}({\bf x})$ on the even
sub-lattice with the spin operators ${\bf S}({\bf y}={\bf x}+{\bf \nu})
=T_{\nu}{\bf S}(x)T_{\nu}^{-1}$ on the odd sub-lattice.

According to \eqref{psi_m} and \eqref{a5} the application of
the translation operator $T_{\nu}$ yields:

\begin{eqnarray}
 & & \sum_{{\bf \nu}}\langle \Psi^{(-)}(\sigma)|T_{{\bf \nu}}S_3({\bf x})
T_{{\bf \nu}}^{-1}|\Psi^{(+)}\rangle\quad\quad\quad
\nonumber\\
 & = & 4\cdot \langle \Psi^{(+)}(\sigma)|S_3({\bf x})|\Psi^{(+)}(\sigma)\rangle\\
 & = & 4\cdot\frac{1}{N/2}\cdot\sigma\,,\quad\quad\quad\sigma=s/2\,,
\nonumber\\
 & = & 4M\,.\nonumber
\end{eqnarray}

We end up with the following expression for the ground state energies
in the antiferromagnetic collinear phase $\alpha > \alpha_0(M)$ in the
first order approximation:

\begin{eqnarray}
E(s,{\bf p}_s=(0,0),\alpha,N) & = & \alpha\Big(2E(s/2,{\bf p}_{s/2}=(0,0),N/2)
\nonumber\\
 & & +\frac{1}{\alpha}2M^2\cdot N\Big)
\end{eqnarray}

We compute in the same manner the expectation values of $H_1$ between the
ground state \eqref{a8} for $s$ odd, $\sigma =s/2+1$ even (${\bf p}_s=(0,\pi)$):

\begin{eqnarray}
E(s,{\bf p}_s=(0,\pi),\alpha,N) & = & \alpha\Big(E(\sigma,{\bf p}_{\sigma}=
(0,0),N/2)\nonumber\\
 & & +E(\sigma-1,{\bf p}_{\sigma}=(\pi,\pi),N/2)\nonumber\\
 & & +\frac{1}{\alpha}2M^2 N\Big)\,.
\end{eqnarray}

%***************************************************************************

\section{Second order perturbation theory in the strong frustration
limit}\label{2nd_order}

The evolution of the energy eigenvalues and transition matrix elements
of a Hamiltonian

\begin{eqnarray}\label{hb}
 & & H_2+\beta H_1
\end{eqnarray}
depending on some parameter $\beta$ is described by a closed system of
differential equations \cite{fl98b}. In particular one finds for
the second derivative of the ground state energy:

\begin{eqnarray}\label{evolution}
\frac{d^2E_0}{d\beta^2} & = & -2\sum_n\frac{|T_{n0}|^2}{E_n-E_0}
\end{eqnarray}
where
\begin{eqnarray}
T_{n0} & = & \langle n|H_1|0\rangle\,.
\end{eqnarray}

$|0\rangle$ and $|n\rangle$ denote the ground state and the exited states of
\eqref{hb} with energy eigenvalues $E_0$ and $E_n$, respectively.

We are now going to evaluate \eqref{evolution} in the strong frustration limit
$\beta=1/\alpha\rightarrow\infty$. The ground state of $H_2$ in the
sector with total spin $s$ ($s/2$ even) is given in \eqref{product}. 
%It turns out that in the thermodynamical limit the leading contributions 
%to the transition probabilities $|T_{n0}|^2$ arise from the excited
%states:
In the following we will consider those contributions to the
transition amplitudes, which arise from the excited states:

\begin{eqnarray}
\Psi^*(s,{\bf p}_s=(0,0),N) & = & \quad\quad\quad\quad\quad\quad\quad
\quad\quad\quad\quad\quad\nonumber
\end{eqnarray}
\vspace{-0.7cm}
\begin{eqnarray}\label{ex_st}
\frac{1}{\sqrt{2}}\left(\Psi^{(+)}
(\sigma+1,{\bf p},N/2)\Psi^{(-)}(\sigma-1,-{\bf p},N/2)
\right. & & \nonumber\\
\quad +\left.\Psi^{(+)}(\sigma-1,-{\bf p},N/2)\Psi^{(-)}(\sigma+1,
{\bf p},N/2)\right) & &\label{b4}
\end{eqnarray}
with energy $E^{(+)}(\sigma+1,{\bf p},N/2)+E^{(-)}(\sigma-1,{\bf p},N/2)$
and $\sigma=s/2$. $E^{(\pm)}(\sigma',{\bf p},N/2)$ are the lowest energies
on the subsystems for a given spin $\sigma'$ and momentum ${\bf p}$.
The momenta ${\bf p}$ and $-{\bf p}$ on the subsystems have to add
up to zero in order to produce a translationally invariant
(momentum ${\bf p}_s={\bf 0}$) state on the whole lattice. Similarly
the sublattice spins $\sigma+1$ and $\sigma-1$ add up to the total
spin $s$ of the whole system. Any other distribution of the total
spin $s$ on the sublattices will increase the total energy in
comparison with $E^{(+)}(\sigma+1,{\bf p},N/2)+E^{(-)}(\sigma-1,
{\bf p},N/2)$.

%Note that the momenta ${\bf p}$, $-{\bf p}$ on the subsystems $(+)$
%and $(-)$ have to add up to zero in order to produce a translationally
%invariant -- momentum ${\bf p}_s={\bf 0}$ -- state on the whole lattice.

The computation of the transition matrix elements yields:

\begin{eqnarray}
\langle \Psi^{(+)}_{\sigma+1,{\bf p}}\Psi^{(-)}_{\sigma-1,-{\bf p}}|H_1|\Psi^{(+)}
_{\sigma,{\bf p}_{\sigma}=(0,0)}\Psi^{(-)}_{\sigma,{\bf p}_{\sigma}=(0,0)}\rangle
 & & \nonumber\\[5pt]
=\sum_{{\bf x}\in(+)}\langle \Psi^{(+)}_{\sigma+1,{\bf p}}|S_+(x)|\Psi^{(+)}_
{\sigma,{\bf p}_
{\sigma}=(0,0)}\rangle\quad\quad\quad & &\nonumber\\
\times\sum_{{\bf \nu}}\langle \Psi^{(-)}_{\sigma-1,-{\bf p}}|T_{{\bf \nu}}S_-({\bf x})
T_{{\bf \nu}}^+
|\Psi^{(-)}_{\sigma,{\bf p}_{\sigma}=(0,0)}\rangle & &\nonumber\\
=\,\frac{N}{2}\cdot\left(1+e^{ip_1}+e^{ip_2}+e^{i(p_1+p_2)}\right)
\quad\quad\quad\quad & &\nonumber\\
\times\langle \Psi^{(+)}_{\sigma+1,{\bf p}}|S_+({\bf x}={\bf 0})|
\Psi^{(+)}_{\sigma,{\bf p}=(0,0)}\rangle
\quad\quad\quad & &\nonumber\\
\times\langle \Psi^{(+)}_{\sigma-1,-{\bf p}}|S_-({\bf x}={\bf 0})|
\Psi^{(+)}_{\sigma,{\bf p}=(0,0)}\rangle
\quad\quad\quad & &\,.\label{b5}
\end{eqnarray}

Here we have used the transformation properties \eqref{a5}
of the sublattice eigenstates $\Psi^{(+)}(\sigma\pm 1,{\bf p})$ under
the translation operators $T_{\nu}$.

A similar expression for the transition matrix elements to the second
component in the excited state \eqref{b4}, is obtained by the substitution
${\bf p}\rightarrow -{\bf p}$ in the prefactor on the right-hand side 
of \eqref{b5}.

Combining both results, we obtain for the transition probabilities
to the excited states \eqref{b4}:

\begin{eqnarray}
|T_{n0}|^2 & = & |\langle \Psi^*_{s,{\bf p}_s}|H_1|\Psi^*_{s,{\bf p}_
s}\rangle |^2\quad ;\quad {\bf p}_s=(0,0)\nonumber\\
 & = & 2\left(\frac{N}{2}\right)^2\cdot\left(1+\cos p_1\right)^2\left(1+\cos p_2
\right)^2\nonumber\\
 & & \times |\langle \Psi^{(+)}_{\sigma+1,{\bf p}}|S_+(0)|\Psi^{(+)}_{\sigma,{\bf p}_
{\sigma}=(0,0)}\rangle |^2\nonumber\\
 & & \times |\langle \Psi^{(+)}_{\sigma-1,-{\bf p}}|S_-(0)
|\Psi^{(+)}_{\sigma,{\bf p}_{\sigma}=(0,0)}\rangle |^2\,.\label{b6}
\end{eqnarray}

For the evaluation of the second order term \eqref{evolution} in the
perturbation expansion of \eqref{hb} around $\beta=1/\alpha=0$ we also
need the energy differences

\begin{eqnarray}
E_n-E_0 & = & E_{n_1}({\bf p},\sigma+1,N/2)+E_{n_2}(-{\bf p},\sigma-1,N/2)
\nonumber\\
 & & -2E_0({\bf p}_0={\bf 0},\sigma,N/2)\,.
\end{eqnarray}

\begin{figure}[]
\centerline{\hspace*{0mm}\epsfig{file=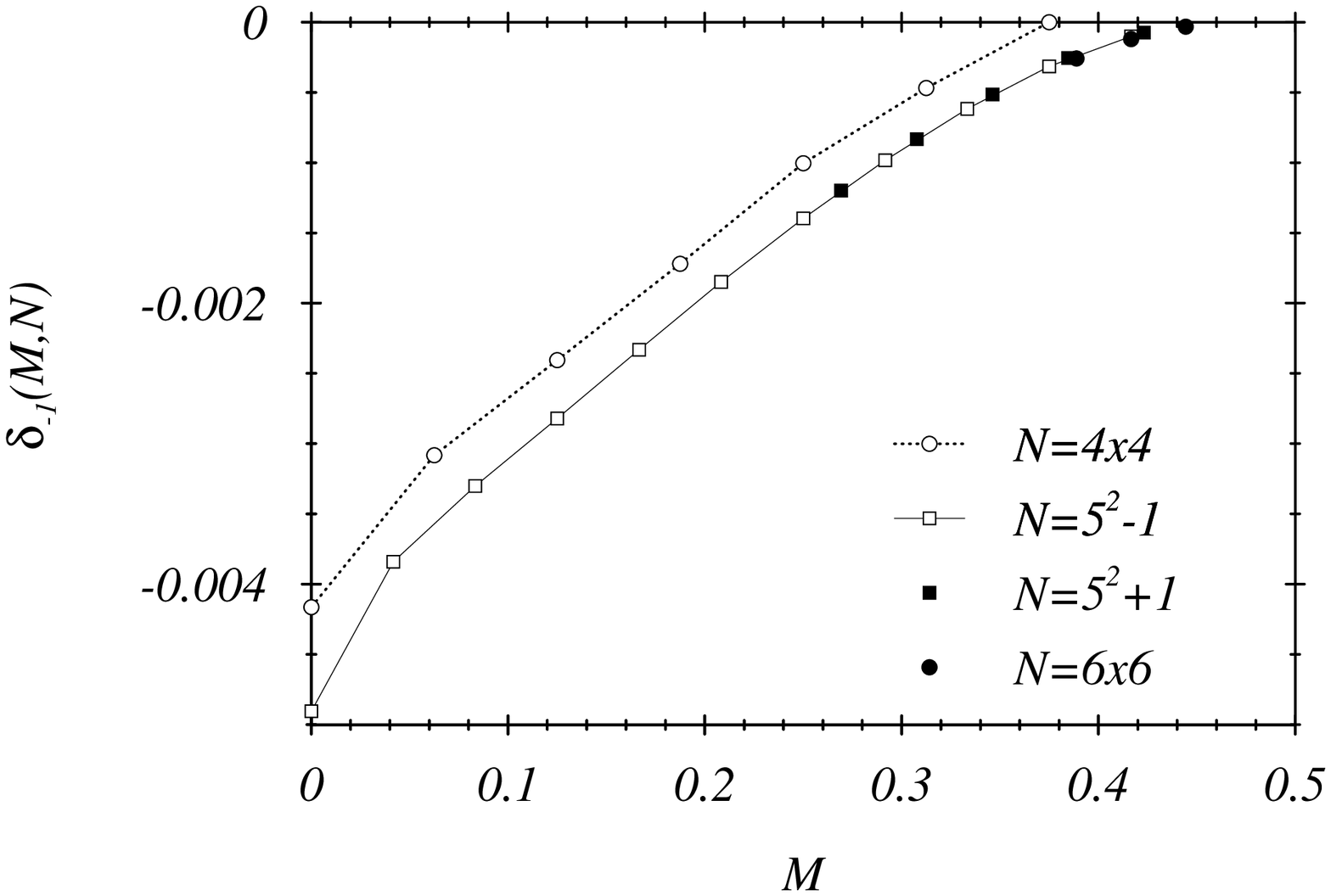,width=7.6cm,angle=0}}
\caption{$M$- and $N$-dependence of the second order contribution
$\delta_{-1}(M)$}
\label{fig10}
\end{figure}

We have computed the energy differences 

\begin{eqnarray}
 & & E_{n_l}({\bf p},\sigma\pm 1,N/2)
-E_0({\bf p}_0={\bf 0},\sigma,N/2)\,,\nonumber\\
 & & \hspace{4.0cm} \sigma=s/2,\,\,\,l=1,2 \nonumber
\end{eqnarray}

on the $\pm$-subsystems
as well as the transition probabilities

\begin{eqnarray}
 & & |\langle \Psi_{n_l,\sigma\pm 1,\pm{\bf p}}^{(+)^*}|S_{\pm}(0)|\Psi_{
\sigma,{\bf p}_{\sigma}={\bf 0}}\rangle|^2
\end{eqnarray}

by means of the recursion method \cite{recursion}. The resulting
coefficient $\delta_{-1}(M)$, which appears in the expansion
\eqref{2nd_gt}:

\begin{eqnarray}
\delta_{-1}(M) & = & -\frac{1}{N}\sum_{n_1, n_2}\sum_{p_1, p_2}
\frac{|T_{n0}|^2}{E_n-E_0}
\end{eqnarray}

is shown in Fig.~\ref{fig10}.

%%%%%%%%%%%%%%%%BEGIN-FIGURE%%%%%%%%%%%%%%%%%
%%%%%%%%%%%%%%%%%END-FIGURE%%%%%%%%%%%%%%%%%%

\end{appendix}
%%%%%%%%%%%%%%%%%%%%%%%%%%%%%%%%%%%%%%%

%%%%%%%%%%%%%%%%%%%%%%%%%%%%%%%%%%%%%%%
%  bibtex
%%%%%%%%%%%%%%%%%%%%%%%%%%%%%%%%%%%%%%%%
%\bibliography{/home/karbach/REFERENCES/references}

\begin{thebibliography}{10}

\bibitem{oshikawa97a}M. Oshikawa, M. Yamanaka, and I. Affleck,
Phys. Rev. Lett. {\bf 78}, 1984 (1997)
%Magnetization plateaus in spin chains: "'Haldane gap"' for half-integer spins
% 1

\bibitem{fl99}
A. Fledderjohann, C. Gerhardt, M. Karbach, K.-H. M\"utter, and R. Wiessner,
Phys. Rev. B{\bf 59}, 991 (1999)
%Soft modes, gaps and plateaus in 1D spin-1/2 antiferromagnetic Heisenberg
%models
% 2

\bibitem{lieb61}E. Lieb, T. Schultz, and D. Mattis, Annals of Phys. {\bf 16},
407 (1961)
%Two soluble models of an antiferromagnetic chain
% 3

\bibitem{nojiri88}H. Nojiri, Y. Tabunaga, and M. Motokawa, J. de Physique {\bf 49},
C8 1459 (1988)
% 4

\bibitem{honecker00}A. Honecker, M. Kaulke, and K. D. Schotte, Eur. Phys. J.
B\,{\bf 15}, 423 (2000)
%A Spin-1/2 Model for CsCuCl_3 in an External Magnetic Field
% 5

\bibitem{shiramura98}W. Shiramura et al., J. Phys. Soc. Jpn. {\bf 67}, 1548 (1998)
%w shiramura, k-i takatsu, b kurniawan, h tanaka, h uekusa, y ohashi, k takizawa,
%h mitamura
%magnetization plateaus in NH_4CuCl_3
% 6

\bibitem{kolezhuk99}A. K. Kolezhuk, Phys. Rev. B\,{\bf 59}, 4181 (1999)
%magnetization plateaus in weakly coupled dimer spin system
% 7

\bibitem{onizuka00}K. Onizuka, H. Kageyama et al., J. Phys. Soc. Jpn. {\bf 69},
1016 (2000)
%1/3 magnetization plateau in SrCu_2(BO_3)_2 -stripe order of excited triplets-
% 8

%\bibitem{kageyama99}H. Kageyama et al., J. Phys. Soc. Jpn. {\bf 68}, 1821 (1999);
%H. Kageyama et al., Phys. Rev. Lett. {\bf 82}, 3168 (1999)
%JPSJ: anomalous magnetizations in single crystalline SrCu_2(BO_3)_2
%PRL: exact dimer ground state and quantized magnetization plateaus in the
%two-dimensional spin system SrCu_2(BO_3)_2
% 

\bibitem{shastry81}B. S. Shastry, B. Sutherland, Physica {\bf 108}\,B, 1308 (1981)
% 9

\bibitem{miyahara99}S. Miyahara, K. Ueda, Phys. Rev. Lett.\,{\bf 82}, 3701 (1999)
%exact dimer ground state of the two dimensional heisenberg spin system
%SrCu_2(BO_3)_2
% 10

\bibitem{muellerhartmann00}E. M\"uller-Hartmann, R. P. Singh, C. Knetter, 
and G. Uhrig,
Phys. Rev. Lett.\,{\bf 84}, 1808 (2000)
% 11

\bibitem{momoi00}T. Momoi, K. Totsuka,
Phys. Rev. B\,{\bf 62}, 15067 (2000);
Y. Fukumoto, A. Oguchi, J. Phys. Soc. Jpn.
{\bf 69}, 1286 (2000); Y. Fukumoto, 
J. Phys. Soc. Jpn. {\bf 70}, 1397 (2001)
%cond-mat/0012396
%magnetization plateaus of the shastry-sutherland model for SrCu_2(BO_3)_2:
%spin-density wave, supersolid, and bound states

%magnetization process in the Shastry-Sutherland system
%SrCu_2(BO_3)_2: results of third-order dimer expansion

%magnetization plateaus in the Shastry-Sutherland model for
%SrCu_2(BO_3)_2: results of fourth-order perturbation
%expansion with a low-density approximation


\bibitem{marshall55}W. Marshall, Proc. Roy. Soc. (London) A\,{\bf 232},
48 (1955)
%antiferromagnetism
% 12

\bibitem{schulz92}H. J. Schulz, T. A. L. Ziman,
Europhys. Lett.\,{\bf 18}, 355 (1992)
%finite-size scaling for the two-dimensional frustrated quantum heisenberg
%antiferromagnet
% 13

\bibitem{zhitomirsky00}M. E. Zhitomirsky, A. Honecker, and O. A. Petrenko,
Phys. Rev. Lett.\,{\bf 85}, 3269 (2000)
%Field induced ordering in highly frustrated antiferromagnets
% 14

\bibitem{lozovik93}Yu. E. Lozovik, O. I. Notych,
Solid St. Commun. {\bf 85}, 873 (1993)

\bibitem{recursion}
A. Fledderjohann, M. Karbach, K.-H. M\"utter, and P. Wielath, J. Phys. C\,{\bf 7},
 8993 (1995);
%Computation of dynamical structure factors with the recursion method
A. Fledderjohann, K.-H. M\"utter, M.-S. Yang, and M. Karbach,
Phys. Rev. B\,{\bf 57}, 956 (1998)
%from one to two dimensions in quantum spin systems
% 15

\bibitem{fabricius91b}K. Fabricius, U. L\"ow, and K.-H- M\"utter,
Phys. Rev. B\,{\bf 44}, 9981 (1991)
%complete solution of the two-dimensional antiferromagnetic heisenberg
%model on small lattices
% 16

\bibitem{yang97}M.-S. Yang, K.-H. M\"utter, Z. Phys. B\,{\bf 104}, 117 (1997)
%the two dimensional antiferromagnetic heisenberg model in the presence
%of an external field
% 17

\bibitem{honecker00b}A. Honecker, preprint (2000)
% 18

\bibitem{hohenberg75}P. C. Hohenberg, W. F. Brinkman,
Phys. Rev. B\,{\bf 10}, 128 (1974)
%sum rules for the frequency spectrum of linear magnetic chains
% 19

\bibitem{mueller82}G. M\"uller,
             Phys. Rev. B \textbf{26}, 1311 (1982)
%sum rules in the dynamics of quantum spin chains
% 20

\bibitem{fl98b}
A. Fledderjohann, K.-H. M\"utter, and M. Karbach, Europ. Phys. J. B\,{\bf 5},
487 (1998)
%The 1D spin-1/2 AF-Heisenberg model in a staggered field
% 18





\end{thebibliography}
%\bibliographystyle{prsty}

\end{document}